\documentclass[submission,copyright,creativecommons]{eptcs}
\providecommand{\event}{SynCoP 2016} 
\usepackage{amsmath,amssymb,mathtools,amsthm}
\usepackage{caption3}
\usepackage{subfig}
\usepackage{graphicx}
\usepackage{epsfig}
\usepackage{algorithm}
\usepackage{algpseudocode}

\def\smallskip{\vskip\smallskipamount}
\def\medskip{\vskip\medskipamount}
\def\bigskip{\vskip\bigskipamount}
\newcounter{thm}[section]

\def\claim#1#2{\par\medskip\noindent\refstepcounter{thm}\hbox{\bf #1 \arabic{section}.\arabic{thm}} (#2).
 \it\ 
}
\def\endclaim{
	\par\medskip}
\newenvironment{thm}{\claim}{\endclaim}
\newcommand{\plant}[1]{$\mathcal{#1}$}
\DeclareMathOperator*{\argmax}{arg\,max}
\DeclareMathOperator*{\argmin}{arg\,min}
\makeatletter
\def\verbatim@font{\linespread{0.5}\normalfont\ttfamily}
\makeatother

\title{Rate Reduction for State-labelled Markov Chains\\ with Upper Time-bounded CSL Requirements}
\author{Bharath Siva Kumar Tati \qquad\qquad Markus Siegle
\institute{Universit\"at der Bundeswehr M\"unchen\\ Germany}
\email{Bharath.Tati@unibw.de \qquad\qquad Markus.Siegle@unibw.de}
}
\def\titlerunning{Rate Reduction for SMCs with Upper Time-bounded CSL Requirements}
\def\authorrunning{B. S. K. Tati \& M. Siegle}
\begin{document}
\maketitle

\begin{abstract}
This paper presents algorithms for identifying and reducing a dedicated 
set of controllable transition rates of a
state-labelled continuous-time Markov chain model.
The purpose of the reduction is to make states to satisfy a given 
requirement, specified as a CSL upper time-bounded Until formula.
We distinguish two different cases, depending on the type of probability bound. 
A natural partitioning of the state space allows us to develop possible 
solutions, leading to simple algorithms for both cases.
\end{abstract}
\section{Introduction}

In a production plant, there can be the requirement that
``once started, the production process should be completed within 1 hour
in 95\% of all cases",
or that ``the probability of an alarm during the first
30 min of operation should be at most 5\%".
If the system is modelled as a state-labelled continuous-time Markov chain
(SMC) and the requirements are formulated with the help of continuous
stochastic logic (CSL), they can be verified automatically by
stochastic model checking \cite{Baier2003},
supported by efficient tools such
as the probabilistic model checker PRISM \cite{Kwiatkowska2011} or MRMC \cite{Katoen2011}.

This paper addresses the question of how to improve a given system
(also called plant \plant{P}) when it has been found that \plant{P} violates
such a given formal requirement.
In an earlier paper \cite{Bharath2015} we presented solutions for the case
of \emph{untimed} probabilistic requirements, and building on that work
we now present solutions for the case of \emph{upper time-bounded} \emph{Until}-type
requirements (without nested or multiple \emph{Until} operators).
In general, one could think of many ways of how to modify a system
in order to make it satisfy a given requirement.
We decided to restrict ourselves to \emph{rate reduction}, which
means that some of the system's transition rates may be reduced, but
the structure of the system remains untouched.
We only allow rate \emph{reductions},
as opposed to increasing any rates, motivated by the fact that it is
usually easily possible to slow down some controllable technical process
(machine, processor, etc.), while it may not be possible to accelerate it. The algorithms presented in this paper assume that all rates of the SMC 
are controllable, i.e.\ can be reduced, but of course we are aware of 
the fact that in real life some rates may be out of the control of the 
plant's operator, for instance the rate of external arrivals or of 
failures caused by the environment.
We work with a partitioning of the state space into classes, based on 
the requirement at hand.
Our strategy then is to reduce all the controllable transition rates between some source
class and target class by a \emph{common reduction factor}.
Depending on the case, different sets of transitions may be reduced by
different reduction factors to achieve the goal.
Before the start of the adaptation process, the value of all reduction factors
is 1, and in the end all reduction factors will be still at most 1, but
strictly greater than 0
(which means that no transitions are completely disabled).
Throughout, our intention is to make as many as possible states
of \plant{P} satisfy the user requirement, but it is not always possible
to make all states satisfying.
This paper develops simple, intuitive algorithms, which are first
motivated by examples.
We show the correctness of the algorithms
(while pointing out the limitations of Algorithm~\ref{algorithm2})
and also analyze their complexity.

Related work: A related topic is model checking of parametric Markov chains,
where reachability probabilities take the form of rational functions
\cite{Daws2004,Hahn2010,Hahn2011}.
Here the goal is to find valid parameter values in a multi-dimensional
search space, as described in \cite{Han2008}, where a discretization 
strategy was proposed together with refinement and/or sampling.
Synthesizing optimal rate parameter values is also the goal of 
\cite{Ceska2014}
(in the context of Markov models of biochemical systems),
where time-bounded properties are considered.
Closely related is the so-called \emph{model repair} problem
which occurs when a system violates a given requirement,
which is to be fixed by modifying transition parameters
while at the same time keeping cost at a minimum.
Model repair has been addressed e.g.\ in
\cite{Bartocci2011,Chen2013,Pathak2015},
for parametric DTMC or MDP models, where
solutions are obtained with the help of nonlinear optimization,
sampling/refinement or greedy strategies.
Our approach described here is different from all of the above, since we 
work
with CTMC models which are a priori not parametric, but come with fixed 
rates.
Once some requirement is violated, we seek to identify sets of controllable transitions
and reduce their rates by a common reduction factor in order to make
as many states as possible satisfy the requirement.
We do currently not consider the cost of rate reduction,
and we restrict ourselves to simple algorithms which avoid
expensive multi-dimensional parameter searches.

The rest of the paper is structured as follows:
Sec.~\ref{sec:prelim} provides the basic definitions and
recalls an earlier result for untimed requirements.
In Sec.~\ref{sec:example}, an example is elaborated on in order to explain the
idea of rate reduction.
It illustrates the benefits, but also the limitations of the proposed approach.
The general algorithms are discussed in Sec.~\ref{sec:alg},
which also includes their complexity analysis.
Finally, Sec.~\ref{sec:conclu} summarizes the main findings of the paper
and touches on possible future work.
\vspace{-1em}
\section{Preliminaries}\label{sec:prelim}
A State labelled Markov chain (SMC) is defined as follows:
\begin{thm}{Definition}{SMC}\label{def:plantp}
	A SMC $\mathcal{P}$ is a tuple $(S_\mathcal{P}, R_\mathcal{P}, L_\mathcal{P})$ where
	\begin{itemize}
		\item $S_\mathcal{P}$ is a finite set of states
		\item $R_\mathcal{P}:S_\mathcal{P}\times S_\mathcal{P}\mapsto\mathbb{R}_{\geq 0}$, is the transition function (rate matrix)
		\item $L_\mathcal{P}:S_\mathcal{P}\rightarrow 2^{AP}$ is a state labelling function, where $AP$ is a finite set of atomic propositions
	\end{itemize}
\end{thm}
This definition does not impose any special structural conditions
(such as irreducibility) on the state graph of the Markov chain.
A finite timed \emph{path} $\sigma$ in a SMC \plant{P} is a finite sequence $\sigma = [(s_0,t_0), (s_1,t_1),\allowbreak \cdots,(s_{n-1},t_{n-1}), s_n] \in (S_\mathcal{P} \times \mathbb{R}_{>0})^*\times S_\mathcal{P}$, and with $Paths(s)$ we denote the set of all finite paths originating from state $s$.
Probabilities are assigned to sets of finite timed paths by the usual cylinder set construction on sets of infinite timed paths.
In order to specify user requirements and characterize execution paths of SMCs,
we use a subset of CSL (continuous stochastic logic) \cite{Baier2003}. 

\begin{thm}{Definition}{CSL with upper time bound}\label{def:grammar}
	The grammar for CSL state formulas $\Phi$, $\Phi'$ and path formulas $\varphi$ is given as:
	\begin{eqnarray*}
		\Phi\; \Coloneqq \; q\:|\: \neg\Phi\:|\: \Phi\vee\Phi\:|\:P_{\sim b}(\varphi),\qquad
		\varphi\;  \Coloneqq \; \Phi'\: \mathcal{U}^{\leq t}\: \Phi',\qquad
		\Phi'\;  \Coloneqq \; q\:|\: \neg\Phi'\:|\: \Phi'\vee\Phi' 
	\end{eqnarray*}
\end{thm}
In the definition, $q\in AP$ is an atomic proposition, $\neg$ denotes negation, $\vee$ denotes disjunction, $b\in(0, 1)$
is a probability value, and $\sim\;\in\{\leq ,\geq\}$ a comparison operator. 
$P_{\sim b}(\varphi)$ asserts that the probability measure of the set of paths satisfying $\varphi$ meets the bound given by $\sim b$.
The path formula $\varphi$ is constructed using the
\emph{Until} ($\mathcal{U}$) operator and an upper time bound $t>0$.
Note that we do not consider CSL requirements with nested \emph{Until}
operators (that's why we distinguish between $\Phi$ and $\Phi'$), since parameter adaptations for \emph{Until} operators at 
different levels are interdependent in a complex way. For similar reasons we do not consider requirements with multiple \emph{Until} operators (although the grammar in Def.~\ref{def:grammar} does not exclude them).
We also do not consider the CSL \emph{next} operator since it would be rather
trivial to handle.
\paragraph*{Remark 1}
This paper considers probability bounds $b\in(0,1)$ instead of $b\in[0,1]$,
since the approach presented here does not aim to turn a non-zero probability into zero, or to turn a probability smaller than one into one. So, we do not treat requirements of the form
$P_{\leq 0} (\varphi)$ or $P_{\geq 1}(\varphi)$,
and for similar reasons we also do not treat requirements of the form
$P_{> 0} (\varphi)$ or $P_{< 1}(\varphi)$. Furthermore, there is no need to distinguish between $< b$ and $\leq b$ (or $> b$ and $\geq b$), because in the continuous-time setting the probabilities are the same.

\begin{figure}[t]
	\centering
	\includegraphics[width=0.6\linewidth]{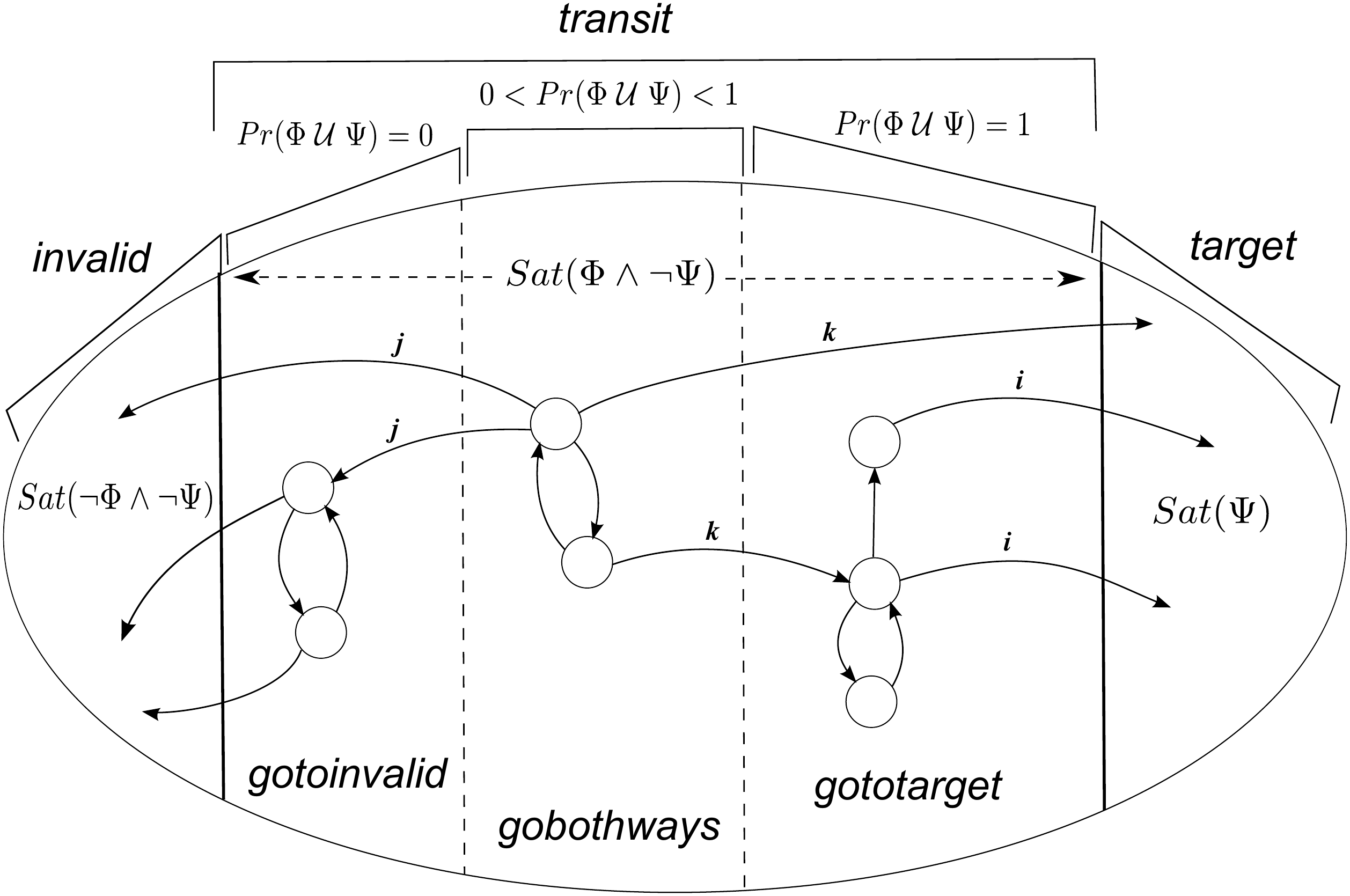}
	\caption{Partitioning of the state space}
	\label{fig:extended_search}
\end{figure}

\begin{thm}{Definition}{Semantics of time-bounded Until path formula}
	The satisfaction relation $\models$ for time-bounded \emph{Until} path formulas is defined as in \cite{Baier2003}:
	\begin{equation*}
		\sigma\models\Phi\:\mathcal{U}^{\leq t}\:\Psi \quad
		\text{if}\quad \exists t' \in [0,t].(\sigma@t' \models \Psi \wedge \forall t'' \in [0,t'). \sigma@t'' \models \Phi)
	\end{equation*}
	where $\sigma@t$ denotes the state occupied by the path $\sigma$ at time $t$.
\end{thm}
The untimed \emph{Until} formula is obtained as: $\Phi\; \mathcal{U}\; \Psi = \Phi\; \mathcal{U}^{< \infty}\; \Psi$. 
Let $Sat(\Phi)$ denote the set of states satisfying state formula $\Phi$,
and let $Pr(s, \varphi) = Pr(\{\sigma \in Paths(s)\; |\; \sigma \models \varphi\})$
denote the probability of the set of $\varphi$-satisfying paths
originating in state $s$.
In order to accomplish the process of rate reduction, we will use
a partitioning of the SMC state space:

\begin{thm}{Definition}{Partitioning of SMC}\label{def:partition}
	Given an SMC \plant{P} and a CSL requirement $\Phi_t = P_{\sim b}(\Phi \; \mathcal{U}^{\leq t} \; \Psi)$.
	Let $\varphi = \Phi \; \mathcal{U}\;  \Psi$, the untimed version of the path formula.
	Then, states belonging to
	\begin{itemize}
		\item $Sat(\neg \Phi \wedge \neg \Psi)$ are placed in class \emph{invalid}
		\item $Sat(\Psi)$ are placed in class \emph{target}
		\item $Sat(\Phi \wedge \neg \Psi)$ are placed in class \emph{transit}
	\end{itemize}
	The \textit{transit} class is further partitioned as:
	\begin{itemize}
		\item $Pr (s, \varphi) = 1$ are placed in class \emph{gototarget}
		\item $Pr (s, \varphi) = 0$ are placed in class \emph{gotoinvalid}
		\item $0 < Pr (s, \varphi) < 1$ are placed in class \emph{gobothways}
	\end{itemize}
\end{thm}
This partitioning is illustrated in Fig.~\ref{fig:extended_search}\footnote{For the purpose of this paper, there is no need to distinguish between classes \emph{invalid} and \emph{gotoinvalid}, but we prefer to separate them for reasons of symmetry.}.
Starting from states of class \textit{gototarget}, the Markov chain 
will eventually reach the \textit{target} class via $\Phi$-states
(almost surely within finite time).
Conversely, states of class \textit{gotoinvalid} do not possess any
path satisfying the given (untimed, and therefore also time-bounded)
\emph{Until} requirement.
From states of class \textit{gobothways}, both of these behaviours
are possible.
During the parameter adaptation procedure, our attention will be on the states of the \textit{transit} class. 
The Partitioning of the state space can be obtained efficiently,
based on the state labelling and by applying standard graph algorithms
\cite{Mehlhorn1984}.
Note that the time bound $\leq t$ and the probability bound $\sim b$ in the given CSL formula have no influence on the partitioning.

As a simple but important fact we emphasize
that the probability of a state satisfying a time-bounded \emph{Until} property will always be less than or equal the probability of that state satisfying the corresponding \emph{untimed} property, i.e.\
\begin{equation}
	\centering
	\forall s. \; \forall t. \; (Pr(s, \Phi\; \mathcal{U}^{\leq t}\; \Psi) \leq Pr (s, \Phi\; \mathcal{U}\; \Psi))
	\label{eq:bounds}
\end{equation}
Furthermore, for an SMC \plant{P}, a subset $X \subseteq S$ of its state set, and a timed or untimed CSL state formula $\Phi$, we introduce the following notation:
$(\mathcal{P}\models_{X}\Phi) \Longleftrightarrow (\forall s\in X:s\models\Phi)$.
As an example, $\mathcal{P} \models_{gobothways} \Phi$ means that all states
from class $gobothways$ satisfy the requirement $\Phi$.

For the purpose of rate reduction, we construct a reduced (parametric, see below) SMC \plant{G} from SMC \plant{P}, where certain transition rates are reduced by factors \emph{i}, \emph{j} and \emph{k} and all the states in classes \emph{invalid} and \emph{target} are made absorbing.

\begin{thm}{Definition}{Reduced parametric SMC \plant{G}}\label{def:plantg}
	Given SMC \plant{P} as in Def.~\ref{def:plantp}, a CSL requirement
	$\Phi_t = P_{\sim b}(\Phi \; \mathcal{U}^{\leq t} \; \Psi)$ and three reduction factors (parameters) $0<i,j,k \leq 1$.
	The reduced SMC \plant{G}$(\Phi, \Psi)$ is defined as a tuple $(S_\mathcal{G},R_\mathcal{G},L_\mathcal{G})$ where: 	
	\begin{itemize}
		\item $S_\mathcal{G} = S_\mathcal{P}$, and the state space partitioning into classes is taken from \plant{P}
		\item $R_\mathcal{G}(s,s') = \begin{cases}
		0 &  s\in target \vee s\in invalid \\
		i\cdot R_\mathcal{P}(s,s') & s\in gototarget \wedge s'\in target \\
		j\cdot R_\mathcal{P}(s,s') & s\in gobothways \wedge s' \in (gotoinvalid \cup invalid) \\
		k\cdot R_\mathcal{P}(s,s') & s\in gobothways \wedge s' \in (gototarget \cup target) \\
		R_\mathcal{P}(s,s') & otherwise
		\end{cases}$
		\item $L_\mathcal{G} = L_\mathcal{P}$
	\end{itemize}
	Let $T_i$ denote the set of transitions whose rate is multiplied by reduction factor $i$ (and analogously for $T_j$ and $T_k$).
\end{thm}
Considering the reduction factors \emph{i}, \emph{j} and \emph{k} as variables, the reduced SMC \plant{G} is a parametric SMC. Once they are fixed, \plant{G} is a standard SMC. Fig.~\ref{fig:extended_search} shows how the transition rates between certain state classes will be multiplied by the reduction factors. The purpose of this multiplication will be explained in the course of the paper.

\subsection{A result for untimed CSL}\label{subsec:untimedresult}

In \cite{Bharath2015}, we presented algorithms for the rate reduction problem
for untimed \emph{Until} requirements, based on the following principle:
In case of an upper probability bound ($\leq b$),
all transition rates from \emph{gobothways} to \emph{gototarget} and to \emph{target}
are reduced by a global factor of $0<k\leq 1$.
Similarly, in case of a lower probability bound ($\geq b$),
all rates from \emph{gobothways} to \emph{gotoinvalid} and to \emph{invalid}
are reduced by a factor of $0<j\leq 1$.
The idea is to thereby influence the branching probabilities of states from
class \emph{gobothways} in the desired direction.
Theorem~3.2 of \cite{Bharath2015} states that following this strategy,
the parameter synthesis problem
for untimed \emph{Until} requirements
can always be solved for all states of class \emph{gobothways}.
We are going to use this result in the sequel, in combination with
Eq.~\eqref{eq:bounds}.
%
\vspace{-1em}
\subsection{Binary Search Method (BSM)}\label{sec:bsm}
In the time-bounded CSL scenario, we use the Binary Search Method (BSM) to approximate the maximum satisfying value of the reduction factor, where we search the range between 0 and 1 up to a precision of a given $\epsilon>0$.
Since a closed-form solution for the transient
state probabilities for a parametric system is not available, 
we use an approximation via BSM with multiple evaluations using uniformisation.
At the start of the reduction process,
the search interval is $(0,1]$.
BSM continues to halve the search interval until its width is at most
the predefined precision $\epsilon$.
In pseudo code, BSM reads as follows, the initial call being
{\tt BSM(0,1)}:
\begin{verbatim}
 BSM(lower, upper)
    {
        while upper-lower > epsilon
         {
            middle = (lower + upper) / 2;
            if "middle satisfies the requirement" then
               lower = middle;
            else
               upper = middle; 
         }
        return lower;	 
    }
\end{verbatim}
As a result, BSM returns the lower bound of the final search interval, where the search is considered unsuccessful if that returned value is zero.
\vspace{-1em}
\section{Example}\label{sec:example}
\begin{figure}[t!]
	\centering
	\includegraphics[width=0.7\linewidth]{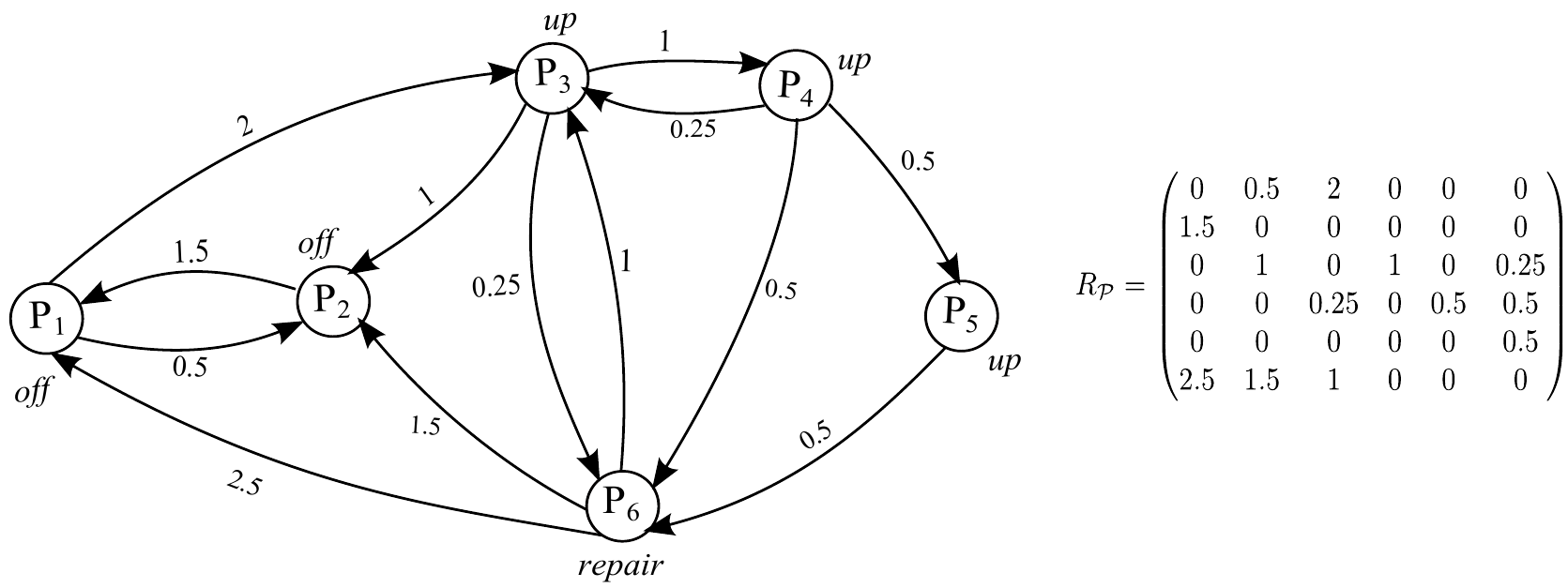}
	\caption{Plant \plant{P} along with rate matrix $R_{\mathcal{P}}$}
	\label{fig:mach_ex_p}
\end{figure}

\begin{figure}[t]
	\centering
	\includegraphics[width=0.8\linewidth]{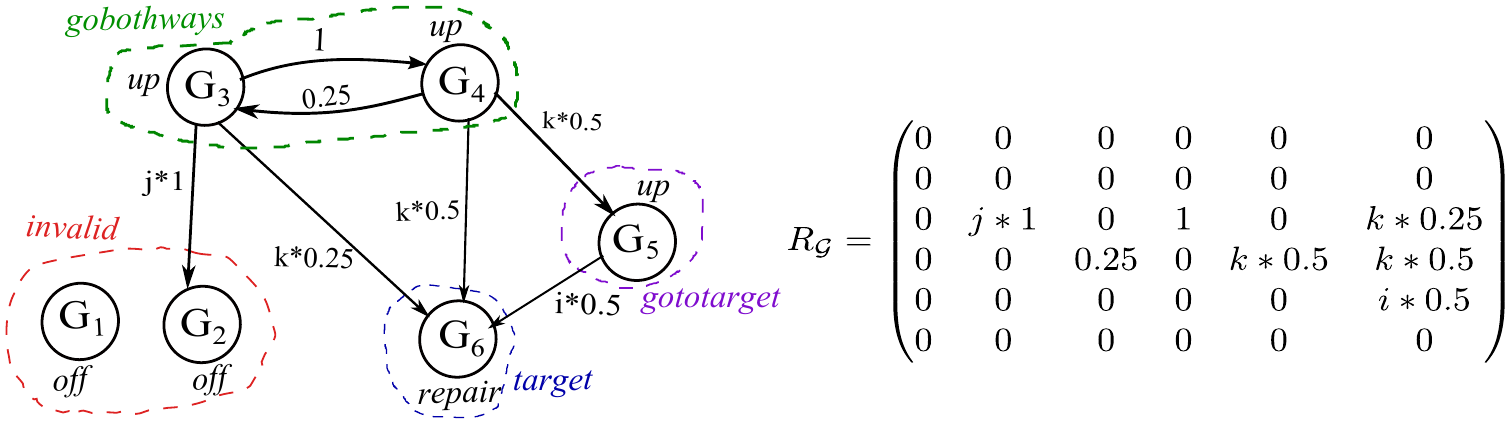}
	\caption{Reduced SMC \plant{G} and its rate matrix}
	\label{fig:mach_ex_g}
\end{figure}

Before giving the general rate reduction algorithms, we will explain our method with the help of an example for the two cases of upper time-bounded CSL \emph{Until} requirements for the model SMC \plant{P} given in Fig.~\ref{fig:mach_ex_p}. Plant \plant{P} (SMC) models an abstract machine with 6 states, which can be \emph{off}, \emph{up} or under \emph{repair}. When the machine is \emph{up}, it can go into \emph{off} or \emph{repair} with some predefined rates. We present our new heuristics/algorithms, partly based on earlier heuristics given in our paper \cite{Bharath2015}, to find reduction factors \emph{i}, \emph{j} and \emph{k} (in case $\Phi = P_{\sim b}(up\; \mathcal{U}^{\leq t}\; repair)$ is violated). In order to create the reduced parametric SMC \plant{G}, we need to partition \plant{P} by using the CSL path formula $\varphi = up\; \mathcal{U}\; repair$ according to Def.~\ref{def:partition}. SMC \plant{G} for this example is given in Fig.~\ref{fig:mach_ex_g}, which also shows the partitioning. Note that for this example, class \emph{gotoinvalid} is empty.

\subsection{Case 1}
We have an abstract model as shown in \plant{P} and the user property $\Phi_1$ in this
case shall be given as in Eq.~\eqref{eq:case1ex}.
\begin{equation}
	\Phi_1 := P_{\leq 0.2}(\varphi), \text{where}\; \varphi = up\; \mathcal{U}^{\leq 5}\; repair
	\label{eq:case1ex}
\end{equation}
Model checking with PRISM \cite{Kwiatkowska2011}, we get the probabilities for each state as follows:
\begin{gather}
	Pr(P_1, \varphi) =0 \leq 0.2 \label{eq:start}\\
	Pr(P_2, \varphi) =0 \leq 0.2\\
	Pr(P_3, \varphi) =0.47323 > 0.2\\
	Pr(P_4, \varphi) =0.83443 > 0.2\\
	Pr(P_5, \varphi) =0.91791 > 0.2\\
	Pr(P_6, \varphi) =1 > 0.2 \label{eq:end}
\end{gather}
From the above probability values, we can see that $Sat(\Phi_1) = \{P_1, P_2\}$. Our focus is only on classes \emph{gobothways} and \emph{gototarget}, as the states from the other classes trivially satisfy/violate the requirement $\Phi_1$. The user property $\Phi_1$ is violated for all the states inside our classes of interest, i.e., $P_3, P_4$ and $P_5$. Hence, the process of rate reduction is required. Before the start of adaptation procedure, the values of \emph{i}, \emph{k} and \emph{j} are equal to 1. Now, since the probability of reaching the \emph{target} class within the specified time bound needs to be reduced, we have to slow down the rates going towards class \emph{target}, i.e., adapt reduction factors \emph{i} and \emph{k}, such that the probabilities of the states in classes \emph{gototarget} and \emph{gobothways} to satisfy $\varphi$ will fall below 0.2.

\begin{figure}[t]
	\centering
	\subfloat[]
	{\label{fig:case1_i}
		\includegraphics[width=.46\linewidth]{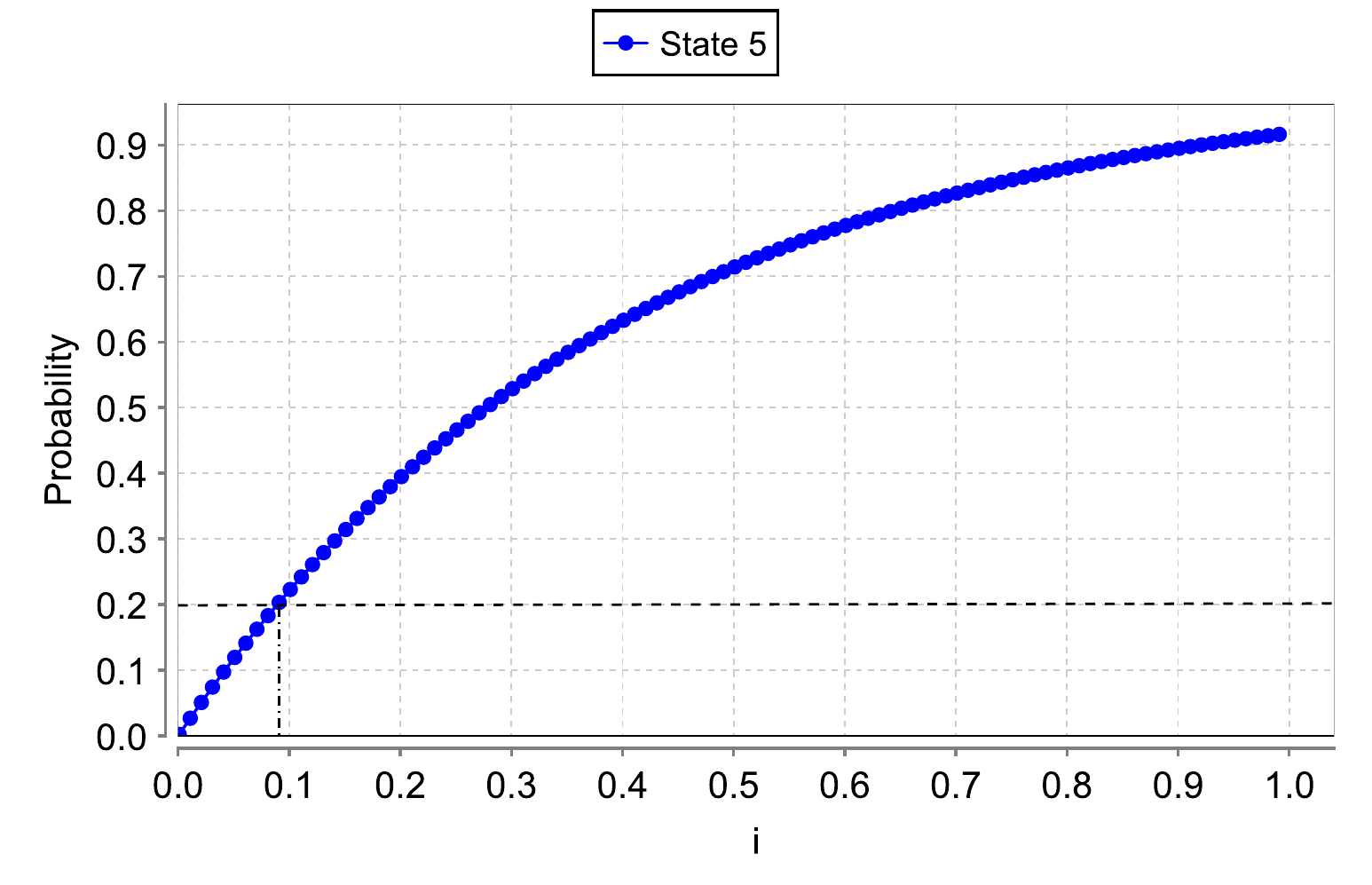}} \quad
	\subfloat[]
	{\label{fig:case1_k}
		\includegraphics[width=.46\linewidth]{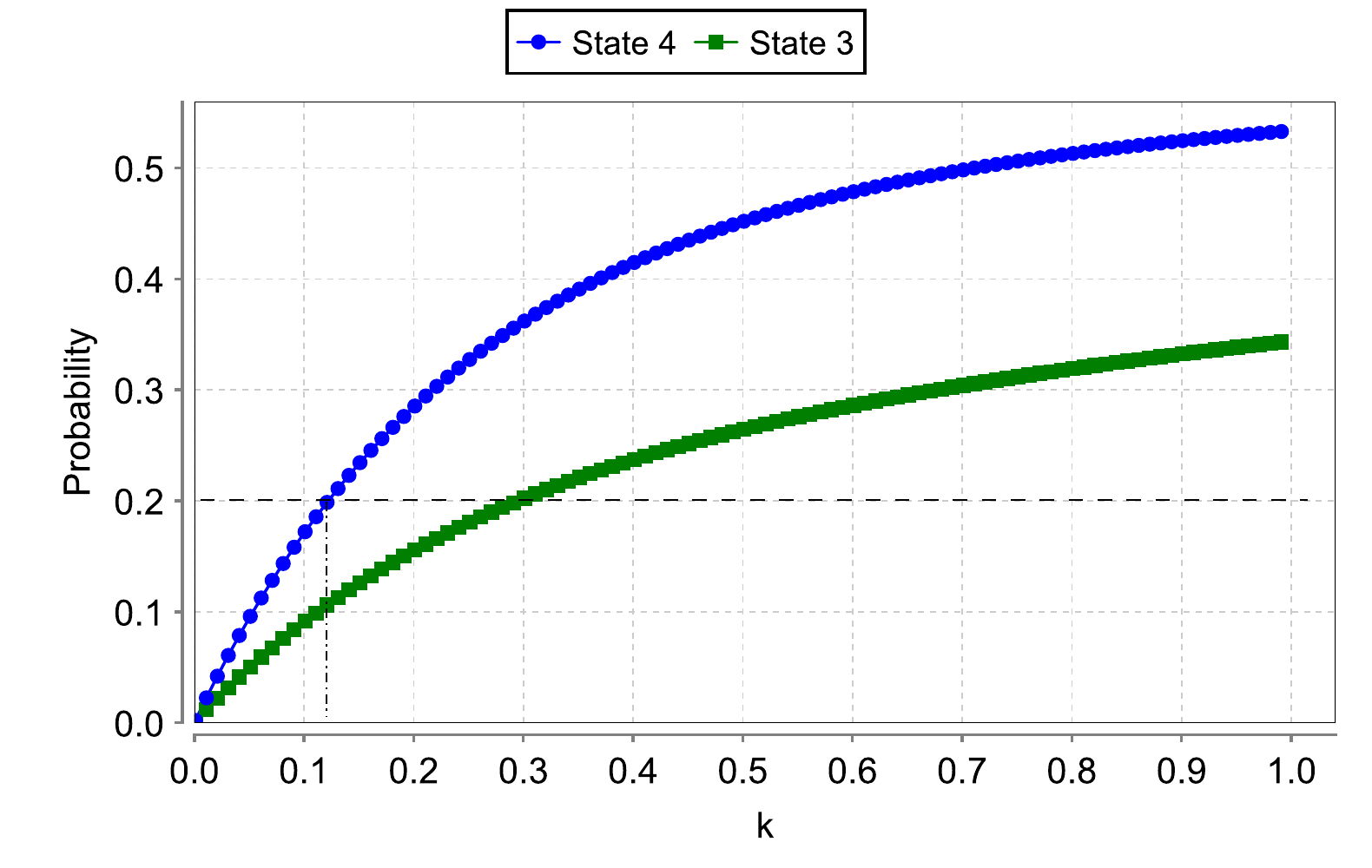}}
	\caption[]{(a) For Case 1, Adapting $i$ for State 5; y-axis is $Pr(P_5, \varphi)$, \\
		(b) For Case 1, Adapting $k$ for states 3 and 4, for fixed $i=0.089$;
		y-axis is $Pr(\star, \varphi)$ }
\end{figure}

Initially, we adapt the \emph{i} value, because the probability of the states in \emph{gototarget} class will not get affected by adapting the \emph{k} value, whereas the vice versa is not true. To graphically show the process of finding the satisfying range of the reduction factor \emph{i}, we plotted\footnote{All the graphs are created using PRISM tool \cite{Kwiatkowska2011}.} the probabilities at equidistant discrete points of \emph{i}. The curve for state $P_5$ of \emph{gototarget} class is shown in the Fig.~\ref{fig:case1_i}.

Finding the reduction factors by solving a set of equations as in the case of untimed CSL is not feasible\footnote{The transient probabilities for a fixed value of the reduction factor are computed numerically via uniformisation, a parametric closed-form solution is not available.} for the time-bounded \emph{Until} case. Hence, we used BSM to approximately find the satisfaction range of \emph{i}, which is $0 < i \leq 0.089$. 

Now, we will focus on the states from class \emph{gobothways} (states $P_3$ and $P_4$). In order to make the states from this class satisfying, we need to adapt the \emph{k} value. In case 1, BSM considers the state with highest probability (here it is state 4) within the class. While adapting \emph{k}, we keep the \emph{i} value fixed to its maximum from the range which we found earlier. By doing so, we obtain the range of \emph{k} to satisfy Eq.~\eqref{eq:case1ex} to be $(0 < k \leq 0.122)$. The graph in Fig.~\ref{fig:case1_k}, shows the probability plots for states 3 and 4 of \plant{P} while reducing factor \emph{k}. The \emph{k} value can also be read from the graph in Fig.~\ref{fig:case1_k}, the curve for state 4 (blue curve) falls below the required probability bound 0.2 at $k=0.122$. Hence, such a \emph{k} range will be the solution for the whole \plant{P}. After applying the reduction factors \emph{i} and \emph{k}, the probabilities of the states in \emph{gobothways} and \emph{gototarget} class modify as follows, and therefore satisfy the user requirement. 
\begin{gather*}
	Pr(P_3, \varphi) =0.10675 \leq 0.2,\qquad
	Pr(P_4, \varphi) =0.19988 \leq 0.2,\qquad
	Pr(P_5, \varphi) =0.19948 \leq 0.2
\end{gather*}
From Eq.~\eqref{eq:bounds}, we know that the probability for a time-bounded user requirement is always less than or equal to the probability of untimed user requirement. Therefore, since the untimed problem can be always solved (see Sec.~\ref{subsec:untimedresult}) we can conclude that the time-bounded variant can also always be solved.

\subsection{Case 2} \label{sec:case2ex}
The user requirement in case 2 shall be as in Eq.~\eqref{eq:case2ex}.
\begin{equation}
	\Phi_2 := P_{\geq 0.95}(\varphi), \text{where again}\; \varphi = up\; \mathcal{U}^{\leq 5}\; repair
	\label{eq:case2ex}
\end{equation}
The actual probabilities of each state for $\varphi$ are the same as shown in equations from \eqref{eq:start} to \eqref{eq:end}. We can see that probabilities of state $P_3$ and $P_4$ (and $P_5$) are lower than the required probability bound i.e., $\geq 0.95$, and thus the user requirement is violated for those states. In case 2, our interest will be only on states from \emph{gobothways} class for the following
reasons:
1) All states from classes \emph{invalid} and \emph{gotoinvalid} trivially cannot satisfy the requirement.
2) States from \emph{gototarget} may or may not satisfy the requirement,
but if they don't, nothing can be improved by reducing any rate
(one would have to increase the rate from \emph{gototarget} to \emph{target},
but this is not one of our options). 

\begin{figure}[t]
	\centering
	\subfloat[]
	{\label{fig:case2}
		\includegraphics[width=.45\linewidth]{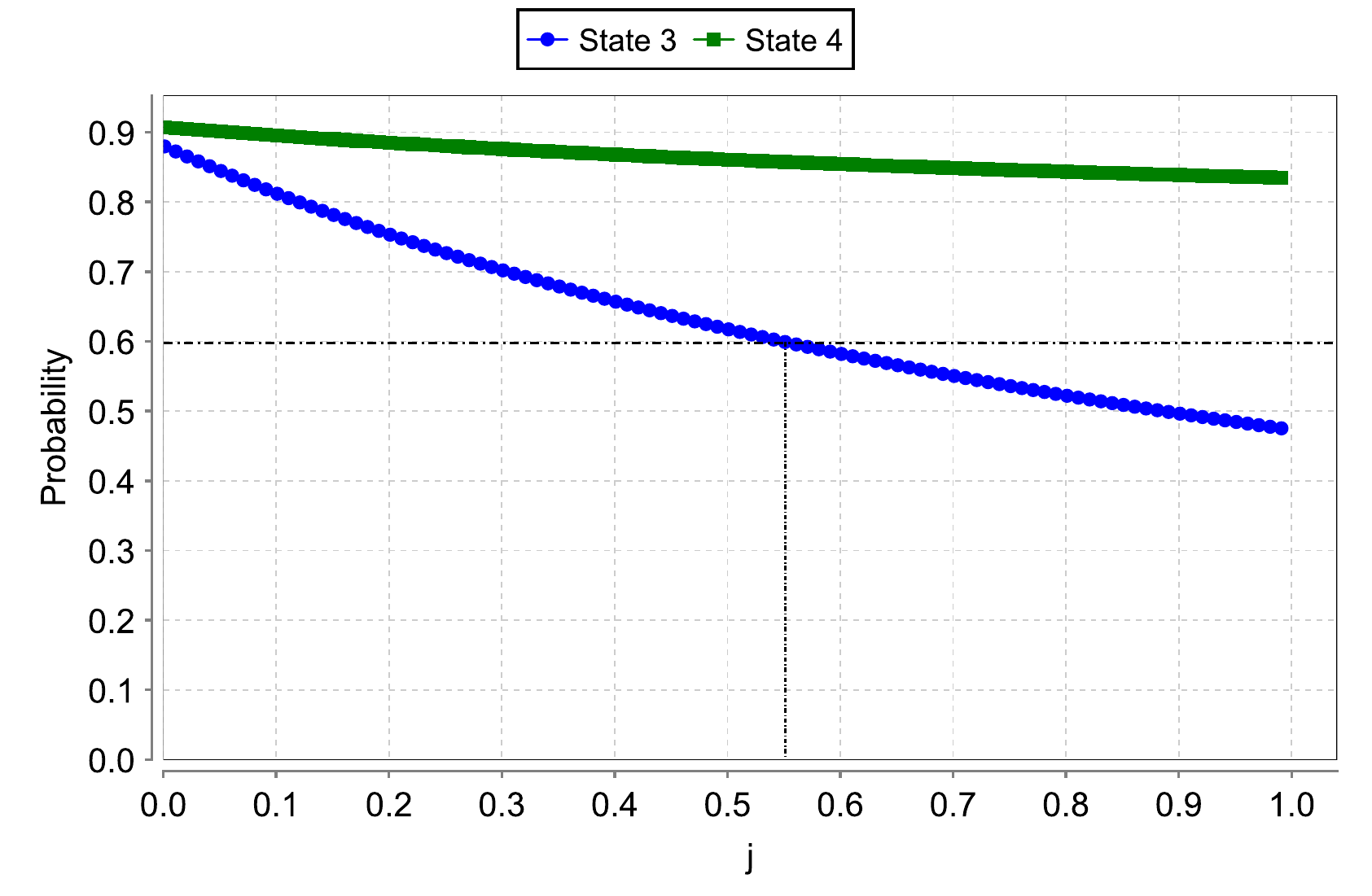}} \quad
	\subfloat[]
	{\label{fig:intersect_k}
		\includegraphics[width=.45\linewidth]{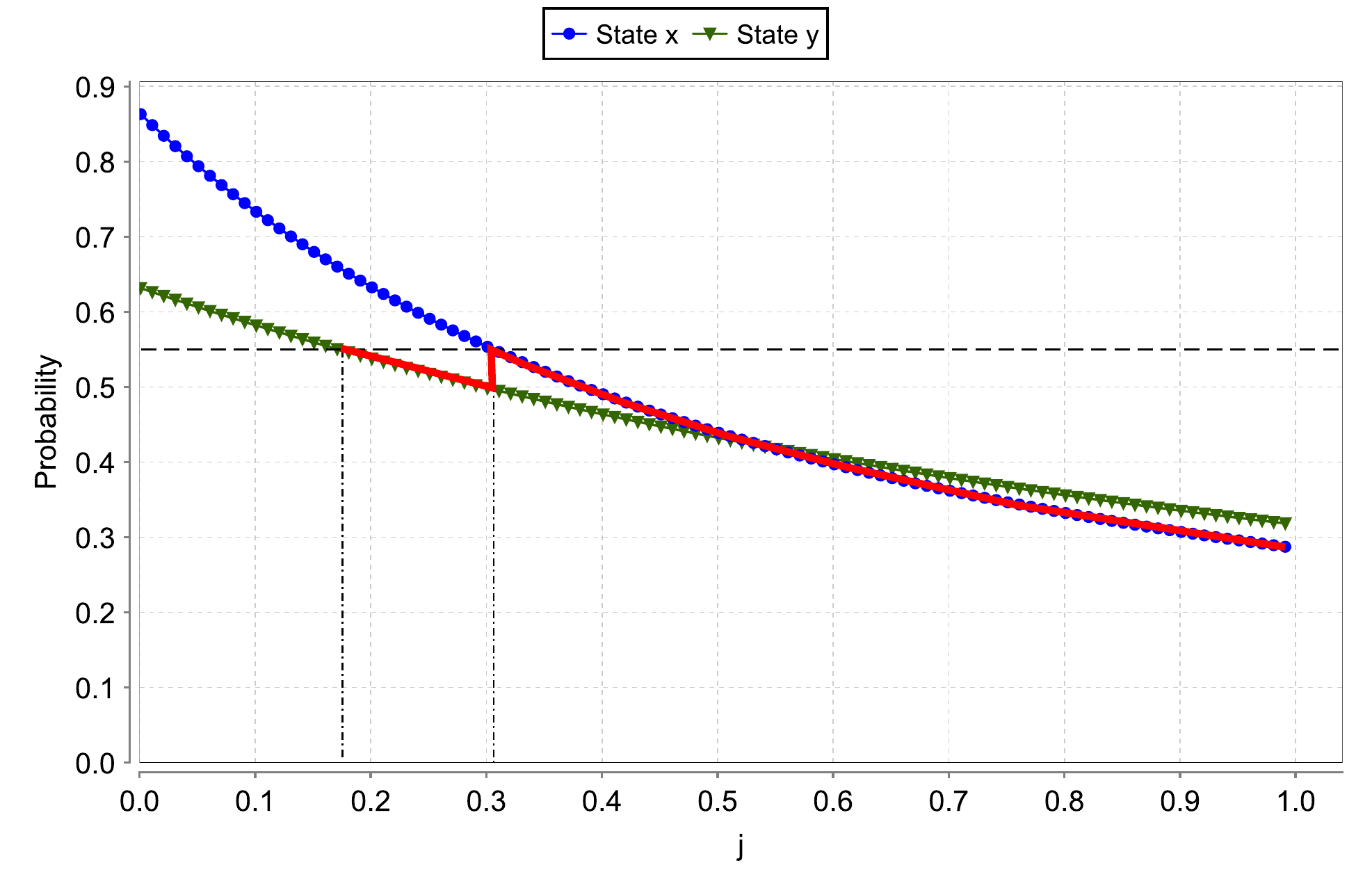}}
	\caption[Graphs]{(a) Adapting \emph{j} for states 3 and 4 for Case 2; y-axis is $Pr(\star, \varphi)$,\\
		(b) Intersecting curves while adapting \emph{j} (for a different example)}
\end{figure}

Here, since the probability should be increased beyond 0.95, the transition rates from \emph{gobothways} to \emph{gotoinvalid} or \emph{invalid} should be reduced, in order to make the transitions between \emph{gobothways} to \emph{target} more likely to happen. For BSM, we have to reduce \emph{j}, and so we choose the state with least probability (here it is state 3) within the \emph{gobothways} class. The graph in Fig.~\ref{fig:case2} shows the probability plots of states 3 and 4 for $\varphi$. As it can be seen, while \emph{j} approaches 0, the probability for $\varphi$ reaches a value of 0.89 (but it does not reach 1.0 due to the time bound), but as per the requirement it should be greater than 0.95. In this case, our algorithm fails to find a satisfying solution.

However, let us take another user requirement for the same case 2 as given in Eq.~\eqref{eq:case2_1ex}, 
\begin{equation}
	\Psi_2 := P_{\geq 0.6}(\varphi), \text{where again}\; \varphi = up\; \mathcal{U}^{\leq 5}\; repair
	\label{eq:case2_1ex}
\end{equation}
Notice the change in the probability bound. Now, from the graph in Fig.~\ref{fig:case2}, we can observe that the satisfying range of \emph{j} to be approximately $0 < j \leq0.548$, where the curve is above required probability bound (i.e., $\geq 0.6$). For the current \emph{j} value, check all the states from \emph{gobothways} class just to make sure they satisfy $\Phi_2$. In general, reducing \emph{j} makes the unbounded requirement $Pr(\Phi_1\:\mathcal{U}\:\Phi_2)$ larger (even reaches prob. 1), but it slows down the process. Hence the existence of a solution in Case 2 depends on the combination of time and probability bounds given in the user requirement.

\section{Algorithms}\label{sec:alg}

This section explains the general rate reduction algorithms for cases 1 and 2 of upper time-bounded CSL \emph{Until} properties. But before we present the algorithms, we need to discuss the important phenomenon of intersecting curves.
In the example from Sec.~\ref{sec:example}, the probability curves for different
states never showed any intersection points.
However, in general the probability curves for different states may 
intersect, as we can observe in other examples.
Fig.~\ref{fig:intersect_k} shows such an example of intersecting curves for varying
reduction factor $j$, but intersection of curves may also
occur when varying factors $i$ or $k$.
For that reason, once the value of the reduction factor has been determined
for the state with the highest (in case 1) or lowest (in case 2) original
probability, one needs to recheck all the other states from the class.
As long as at least one of them violates the requirement, the value of the reduction
factor has to be reduced further.
This is the reason why our algorithms in this section employ while-loops for determining the reduction factors.

\subsection{Case 1}\label{sec:case1}
The CSL requirement in this case is given as $\Phi_1 := P_{\leq b}(\Phi\; \mathcal{U}^{\leq t}\; \Psi)$,
and the state space of \plant{P} is partitioned according to Def.~\ref{def:partition}. For any state $s$ from \emph{invalid} $\cup$ \emph{gotoinvalid},
it holds that $Pr(s, \Phi\; \mathcal{U}^{\leq t}\; \Psi) =0$,
and since $0<b$ those states trivially satisfy the requirement.
Similarly, for states from class \emph{target} the probability
to follow a satisfying path is 1, thus \emph{target}-states trivially violate
the requirement (and nothing can be done about that).

Therefore, the algorithm will only address states from classes \emph{gobothways}
and \emph{gototarget}. Based on the fact stated by Eq.~\eqref{eq:bounds} and the fact that the untimed problem can always be solved (see Sec.~\ref{subsec:untimedresult}), the problem for upper time-bounded \emph{Until} as in $\Phi_1$ can always be solved for these two classes.
The algorithm proceeds by first adapting factor $i$ on transitions from
\emph{gototarget} to \emph{target}. It starts to reduce the probability for $s_h$, the state from class \emph{gototarget} with the highest probability. Since the probability curves of different states within class \emph{gototarget} may intersect with each other, once a satisfying $i$ value is found for $s_h$, we need to check whether all the other states of class \emph{gototarget} also satisfy $\Phi_1$. As long as any of the states violates $\Phi_1$, we continue the process of reducing $i$ (while loop) by selecting the state with highest probability at that instance. Once all states from class \emph{gototarget} satisfy $\Phi_1$ factor $i$ will remain fixed. As a second step, a similar procedure is followed for class \emph{gobothways}, where the algorithm reduces factor $k$ on transitions from
\emph{gobothways} to \emph{gototarget} $\cup$ \emph{target}. Again, because of possible intersection of probability curves, the process needs to be continued for class \emph{gobothways} as motivated above.
Algorithm~\ref{algorithm1} formalizes this rate reduction process for Case 1.
\begin{algorithm} [!t]
	\caption{Rate reduction for $\Phi_1 := P_{\leq b}(\Phi\; \mathcal{U}^{\leq t}\; \Psi)$} \label{algorithm1}
	\begin{algorithmic}
		\State \textbf{Input:} SMC \plant{P} and time-bounded CSL \emph{Until} property $\Phi_1$.
		\State \textbf{Output:} Satisfying \emph{i}, \emph{k} reduction factors and corresponding sets of transitions $T_i, T_k$
		\If {$ gobothways \cup gototarget = \emptyset$}
		\State quit \Comment\textit{No states whose probabilities can be modified}
		\Else
		\If{$\mathcal{P}\models_{gobothways\cup gototarget} \Phi_1$}\Comment\textit{No need for rate reduction}
		\State quit 
		\Else 
		\State $\mathcal{G} = (S_\mathcal{G},R_\mathcal{G},L_\mathcal{G})$; $i=1$; $k=1$  \Comment\textit{Construct reduced SMC (Def.~\ref{def:plantg}), initialize red. factors}
		\While{TRUE} \Comment\textit{ Find $i$ for class $gototarget$}
		\State for current \emph{i}, find $s_h = \argmax_{s}\{Pr(s, \Phi \mathcal{U}^{\leq t} \Psi)\: |\: s \in gototarget\}$
		\State Apply BSM to $s_h$ to find satisfying \emph{i}
		\State Check all $s \in gototarget$ for current $i$
		\If{any of the states still violates $\Phi_1$}
		\State continue
		\Else
		\State fix factor \emph{i} and corresponding set $T_i$ ; break
		\EndIf
		\EndWhile
		\While{TRUE} \Comment\textit{Find $k$ for class $gobothways$}
		\State for current \emph{k} value, find $s_h' = \argmax_{s}\{Pr(s, \Phi \mathcal{U}^{\leq t} \Psi)\: |\: s \in gobothways\}$
		\State Apply BSM  to $s_h'$ to find satisfying \emph{k}
		\State Check all $s \in gobothways$ for current $k$
		\If{any of the states still violates $\Phi_1$}
		\State continue
		\Else
		\State break
		\EndIf
		\EndWhile
		\State return factors $i$, $k$ and corresponding sets $T_i$, $T_k$
		\EndIf
		\EndIf
	\end{algorithmic}
\end{algorithm} 

\subsection{Case 2}
In this case, the general form of CSL property to be checked is given as $\Phi_2 := P_{\geq b}(\Phi\; \mathcal{U}^{\leq t}\; \Psi)$. As already stated in Case 1, 
for any state $s$ from \emph{invalid} $\cup$ \emph{gotoinvalid},
it holds that $Pr(s, \Phi\; \mathcal{U}^{\leq t}\; \Psi) =0$,
and since $0 \not \geq b$ those states trivially violate the requirement
(and nothing can be done about it).
Similarly, for states from class \emph{target}, the probability
to follow a satisfying path is 1, thus \emph{target}-states trivially satisfy
the requirement.
For any state from  class \emph{gototarget}, the probability of satisfying paths
may be above or below the bound $b$.
For states of the former type, the requirement is satisfied, but for
states of the latter type, it is not and nothing can be done about it,
since one would have to
accelerate the paths towards \emph{target}, which is not possible
(since we only allow rate reductions).

Hence, we only focus on the remaining class \emph{gobothways} for rate reduction.
If the given $\Phi_2$ is violated, then the branching probability
(to eventually reach class \emph{target} instead of class \emph{invalid})
is too low, or the speed of moving to class \emph{target} is too slow.
From the result for untimed \emph{Until} (cf.\ Sec.~\ref{subsec:untimedresult}) we know that the branching
probability can be increased to any desired value (arbitrarily close to 1) by the reduction factor $j$.
Following the similar strategy in case of time-bounded \emph{Until} may lead to a solution, but not always (because of the speed being too slow), as demonstrated by example in Sec.~\ref{sec:case2ex}.
So the strategy is to try reducing factor \emph{j}, and return \emph{j} if a satisfying value is found, else return \emph{fail}. Again, since probability curves may intersect, the search for a satisfying value of $j$ needs to be performed in a while loop.
The general algorithm for this case is given in Algorithm \ref{algorithm2}.

\begin{algorithm} [!t]
	\caption{Rate reduction for $\Phi_2 := P_{\geq b}(\Phi\; \mathcal{U}^{\leq t}\; \Psi)$} \label{algorithm2}
	
	\begin{algorithmic}
		\State \textbf{Input:} SMC \plant{P} and time-bounded CSL \emph{Until} property $\Phi_2$.
		\State \textbf{Output:} Satisfying \emph{j} reduction factor and corresponding set of transitions $T_j$, or \emph{fail}
		\If {$ gobothways = \emptyset$}
		\State quit \Comment\textit{No states whose probabilities can be modified}
		\Else
		\If{$\mathcal{P}\models_{gobothways}\Phi_2$}\Comment\textit{No need for rate reduction}
		\State quit 
		\Else 
		\State $\mathcal{G} = (S_\mathcal{G},R_\mathcal{G},L_\mathcal{G})$; $j=1$ \Comment\textit{Construct reduced SMC (Def.~\ref{def:plantg}), initialize red. factor}
		\While{TRUE} \Comment\textit{Find $j$ for class gobothways}
		\State for current \emph{j} value find, $s_l = \argmin_{s}\{Pr(s, \Phi\: \mathcal{U}^{\leq t}\: \Psi)\: |\: s \in gobothways\}$
		\State For state $s_l$ apply BSM to reduce \emph{j}
		\If{solution found} \Comment\textit{i.e., if $j>0$}
		\State Check all $s \in gobothways$ with current \emph{j} value
		\If{any of the states still violates $\Phi_2$}
		\State continue
		\Else 
		\State return factor \emph{j} and corresponding set $T_j$
		\EndIf
		\Else
		\State return \emph{fail}
		\EndIf
		\EndWhile
		\EndIf
		\EndIf
	\end{algorithmic}
\end{algorithm}

\subsection{A comment on optimality of Algorithm~\ref{algorithm2}}\label{subsec:optimal}
As argued in Sec.~\ref{sec:case1}, Algorithm~\ref{algorithm1} always succeeds in making all states of $gobothways \cup gototarget$ satisfying, whereas Algorithm~\ref{algorithm2} does not always find as solution. If Algorithm~\ref{algorithm2} fails, then there does not exist a common reduction
factor $j$ that will make all the states of class $gobothways$ to 
satisfy $\Phi_2$.
In this case, some states from $gobothways$ may indeed satisfy $\Phi_2$,
but not all of them.
Algorithm~\ref{algorithm2} is not always optimal in the following sense:
There exist cases where the algorithm fails but where it would be
possible to use a more general form of rate reduction in order to make
more (or maybe even all) states of $gobothways$ to satisfy $\Phi_2$.
In those cases, it is possible to further increase the probability
of certain states from $gobothways$ to satisfy the path formula
$\Phi\: \mathcal{U}^{\leq t}\: \Psi$
by reducing not only
the rates from class $gobothways$ to class $gotoinvalid \cup invalid$
(as Algorithm~\ref{algorithm2} does),
but also reducing the rates of certain transitions \emph{among} the states of
class $gobothways$ (i.e.\ transitions within $gobothways$).
However, selecting individual transitions among states of class 
$gobothways$ for rate reduction
is a very difficult task for which there are currently no known efficient
algorithms or even heuristics, so this is beyond the scope of this paper.

\subsection{Complexity}
We assume a SMC with $N$ states and $M= O(N^2)$ transitions. The time complexity for model checking time-bounded \emph{Until} is
the same as for CTMC transient solution by uniformisation, which is known to be
$O(M \cdot q \cdot t)$,
where $q$ is the uniformisation rate
and $t$ is the time bound \cite{Baier2003}
(note that the uniformisation rate, which is at least the maximum of the states 
exit rates, actually decreases as we reduce
transitions rates).
For constructing the reduced SMC $\mathcal{G}$ the state classes
need to be determined and the reduction factors need to be inserted,
all of which can be done in time $O(M)$.
Given the desired precision $\epsilon$, each run of BSM requires $O(\log_2 \frac{1}{\epsilon})$
steps.
Therefore, the overall time complexity of Algorithms~\ref{algorithm1} and 
\ref{algorithm2}
is $O(M \cdot q \cdot t \cdot N \cdot \log_2 \frac{1}{\epsilon})$, where the factor $N$ reflects the iterations of the while loop(s).
\vspace{-1em}
\section{Conclusion}\label{sec:conclu}
In this paper, we have considered systems (also called ``plants'') specified as state-labelled Markov chains, and
user requirements given as upper time-bounded CSL formulas
(without multiple or nested \emph{Until} operators).
Whenever a user requirement is violated by some or all states of the plant,
we try to \emph{repair} the plant by reducing some dedicated sets of its transition rates,
such that eventually the user requirement will be satisfied.
We only allow for a slowdown of transition rates
(but without completely disabling any transitions),
since in most practical situations increasing the
transition rates is not feasible.
Upon model checking, some states in the plant's state space may already satisfy the requirement, whereas some others may not.
Some states will have constant probability which cannot be modified by reducing the rates.
Depending on the type of probability bound for the \emph{Until} formula,
we identified two possible cases for which we have devised simple and
intuitive algorithms along with necessary and sufficient conditions for
the solutions to exist.
The algorithms partition the state space into different classes and find the appropriate sets of transitions between those classes whose rates need to be
reduced, as well as the respective reduction factors.

Our ideal goal is to identify the maximum number of states which can be made to satisfy the user requirement.
We have shown that Algorithm~\ref{algorithm1} (for the upper probability bound)
always achieves this goal of maximality,
but as discussed in Sec.~\ref{subsec:optimal}, Algorithm~\ref{algorithm2}
(for the lower probability bound) is not always optimal.

In this paper, we have not addressed lower time-bounded CSL user requirements,
i.e.\ the case of $\Phi = P_{\sim b}(\Phi\; \mathcal{U}^{\geq t}\; \Psi)$.
Based upon the probability bound, lower time-bounded CSL requirements can be
further divided into two categories to which we refer as cases 3 and 4,
and we are planning to elaborate on these cases in a forthcoming paper. 
Another interesting topic for future work would be to consider
time-bounded CSL formulas with multiple or nested \emph{Until} operators, a difficult
problem, as already hinted at in Sec.~\ref{sec:prelim}. Furthermore, there is the interesting open question of
how to repair a given plant at the level of the high-level modelling 
formalism, instead of at the level of the Markov chain.

\bibliographystyle{eptcs}
\bibliography{mybib}
\end{document}